%% file: ETMM11_latex_Template.tex
\documentclass[twocolumn,a4paper,10pt]{article}
\usepackage{amsmath,amsfonts}
\usepackage{epsfig}
\usepackage{times}
\usepackage{epsfig}
\usepackage{color}
\usepackage{xcolor}
\usepackage{url}
\usepackage[small,hang]{caption2}

\definecolor{dgreen}{rgb}{0.29, 0.33, 0.13}

\makeatletter

\newcounter{numbersec}
\renewcommand{\section}[1]{\par\noindent\stepcounter{numbersec}
\par
\vspace{6pt}
\noindent\textbf{\large   \arabic{numbersec} \hspace*{0.3cm} #1 }
\par
\vspace{2pt}
}
\renewcommand{\subsection}[1]{
\par
\vspace{6pt}
\noindent\textbf{#1}
\par
}
\renewcommand{\subsubsection}[1]{%
\par
\vspace{6pt}
\textbf{#1.}
}
\newcommand{\orlu}{\"Orl\"u}

\newcommand{\Abstract}{\par\vspace{6pt}\noindent\textbf{\large Abstract}\par\vspace{2pt}}
\newcommand{\Acknowledgments}{\par\vspace{6pt}\noindent\textbf{\large Acknowledgments }\par\vspace{2pt}}

\newenvironment{References}{
\par\vspace{6pt}\noindent\textbf{\large References}\par\vspace{2pt}
\begin{small}\begin{list}{ }{
\itemsep0mm \parsep0mm\labelsep0mm\leftmargin0mm
}}
{\end{list}\end{small}}

\makeatother

\title{\vspace*{-12mm}
\LARGE \sc \textbf{  
History effects and near-equilibrium in turbulent boundary layers with pressure gradient \\
}}
\author{ \Large \bf \textit{ 
P. Schlatter $^{1,2}$, R. Vinuesa$^{1,2}$, A. Bobke$^{1,2}$ and R. \orlu$^{1}$  }  \\ \\
\bf$^1$Linn\'e FLOW Centre, KTH Mechanics, SE-100 44 Stockholm, Sweden\\
\bf  $^2$Swedish e-Science Research Centre (SeRC), Stockholm, Sweden \\ \\
\underline{\bf pschlatt@mech.kth.se}
}
\date{}

\input{symbollines.tex}

\begin{document}
%%%%%%%%%%%%%%%%%%%%%%%%%%%%%%%%%%%%%%%%%%%%%%%%%%%%%%%%%%%%%%%%%%%%%%%%%%%%%%%%%%%%%%%%%%%%
%
%\multicolumn{2}{c}{ffffffffffffffffffffffffffffffffffffffff}

%\begin{multicols}{2}[\section{Titre numgtggggggggggggggggggggggggggggggggggggggggéroté.}]
%   blabla sur deux colonnes, c'est plus sérieux. C'est le
%   style qui est généralement utilisé pour écrire des
%   articles.
%\end{multicols}

%
%\begin{multicols}{1}
%   ffffffffffffffffffffffffffffffffffffffffffffffffffffffffffffffffffffffffffffffffffffffffff
%\end{multicols}

\maketitle
\thispagestyle{empty}

%\multicolumn{2}{\centering}{textddddddddddddddddde}

%\Title{Reconstruction of turbulent fluctuations for hybrid RANS/LES simulations using a synthetic eddy method}
%********** Insert here the name(s) and address(es) of the author(s).
%\Author{A.~Author, A.~N.~Otherauthor}
%\Author{N. Jarrin, R. Prosser, F. Billard and D. Laurence}
%\Address{School of Mechanical, Aerospace and Civil Engineering, }
%\Address{The University of Manchester, Manchester M60 1QD, UK}
%%\Address{$^+$ EDF R\&D, 6 quai Watier, 78420 Chatou, France}
%\Email{ N.Jarrin@postgrad.manchester.ac.uk,  }
%%\Email{ r.prosser@manchester.ac.uk, dominique.laurence@manchester.ac.uk}

%
%%%%%%%%%%%% Insert here the abstract body.
%
\Abstract

Turbulent boundary layers under adverse pressure gradients are studied using well-resolved large-eddy simulations (LES) with the goal of assessing the influence of the streamwise pressure development. Near-equilibrium boundary layers were identified with the Clauser parameter $\beta= \delta^{*} / \tau_{w} {\rm d}P_{\infty} / {\rm d}x$. 
The pressure gradient is imposed by prescribing the free-stream velocity. In order to fulfill the near-equilibrium conditions, the free-stream velocity has to follow a power-law distribution. The turbulence statistics pertaining to cases with a constant Clauser pressure-gradient parameter $\beta$ were compared with cases with a non-constant pressure distribution at matched $\beta$ and friction Reynolds number $Re_\tau$. It was noticed that the non-constant cases appear to approach far downstream a certain state of the boundary layer, which is uniquely characterised by $\beta$ and $Re_\tau$. The investigations on the flat plate were extended to the flow around a wing section. Comparisons with the flat-plate cases at matched $Re_{\tau}$ and $\beta$ revealed some interesting features: In turbulent boundary layers with strong pressure gradients in the development history the energy-carrying structures in the outer region are strongly enhanced, which can be detected by the pronounced wake in the mean velocity as well as the large second peak in the Reynolds stresses. Furthermore, a scaling law suggested by Kitsios {\it et al.} (2015), proposing the edge velocity and displacement thickness as scaling parameters,  was tested on a constant pressure gradient case. The mean velocity and Reynolds stress profiles were found to be dependent on the downstream development, indicating that their conclusion might be the result of a too short constant pressure gradient region.

%
%%%%%%%%%%%% Body of the article.
%
\section{Introduction} \label{introduction_section}
Turbulent boundary layers (TBLs) subjected to streamwise pressure gradients (PGs) are of great importance in a wide range of industrial applications, including the flow around a wing or inside a diffuser. Despite their relevance, the effects of PGs on the characteristics of wall-bounded turbulence are still elusive. Since the effect of the pressure gradient on the TBL is closely related to its streamwise development, it is important to define the concept of an {\it equilibrium} boundary layer: according to the strict definition by Townsend (1956), this condition requires the mean flow and Reynolds-stress tensor profiles to be independent of the streamwise position $x$, when scaled with appropriate local velocity and length scales. As also shown by Townsend (1956) this condition is only satisfied by the sink flow, although it is possible to define a less restrictive {\it near-equilibrium} condition when the mean velocity defect $U_{\infty}-U$ is self-similar in the outer region, which in any case dominates at high Reynolds numbers (Marusic {\it et al.}, 2010). Townsend (1956) and Mellor and Gibson (1966) showed that these near-equilibrium conditions can be obtained when the free-stream velocity is defined by a power law as $U_\infty=C(x-x_0)^m$, where $C$ is a constant, $x_0$ is a virtual origin and $m$ the power-law exponent. %In particular, \cite{townsend_1956} showed that $m$ has to be larger than $-1/3$ in order to obtain near-equilibrium conditions, which means that all accelerated TBLs subjected to a favorable pressure gradient (FPG), and with $U_\infty$ distributions defined by a power law, exhibit near-equilibrium. Regarding TBLs subjected to adverse pressure gradients (APGs), {\it i.e.}, only the ones with $U_{\infty}$ defined by a power law as defined above, and satisfying $-1/3 < m < 0$ are in near-equilibrium conditions. Note that \cite{townsend_1956} also showed that a freestream velocity distribution defined by an exponential function as $U_{\infty}=C \exp \left (\mu x \right )$ (with $\mu$ being a constant) may also lead to near-equilibrium conditions, although only if $\mu >0$, {\it i.e.}, only in the case of FPG TBLs. 
An additional interesting conclusion is the fact that the widely studied zero pressure gradient (ZPG) TBL, see \emph{e.g.} Schlatter {\it et al.} (2009) or Bailey {\it et al.} (2013), driven by a constant freestream velocity, is a particular case of the general near-equilibrium TBLs proposed by Townsend (1956) and Mellor and Gibson (1966) where $m=0$. Note that it is relatively common in the literature to refer to ``self-similar'' boundary layers, where as discussed above the only case in which complete self-similarity is observed is the sink flow. %For instance, \cite{Skare:1994uf} obtained an experimental APG TBL with a $U_{\infty}$ distribution given by a power law, and with $m=-0.23$, which in principle would lead to near-equilibrium conditions. Nevertheless, they claimed that their boundary layer was self-similar, although they reported variations in the skin friction coefficient of around $\pm 18\%$ over their test section. There is therefore some discrepancy in the terminology and interpretation of PG TBL data, motivated in part by the difficulties arising from setting up canonical PG TBLs, the wide range of parameters used to characterize pressure gradient effects \cite{Monty:2011et}, as well as the great impact of history effects on the local state of the TBL \cite{schlatter2012turbulent}, which leads to additional difficulties in the interpretation of the results.

%\cite{Perry:2002cg} attempted to develop a methodology to compute mean flow, Reynolds stress components and total shear stress profiles, for any given freestream velocity distribution. Although this is a challenging task, especially in the general non-equilibrium case they considered, they reached some success with several empirical closures, and acknowledged the lack of data in the literature to accurately compute such streamwise evolutions. They identified the most relevant parameters determining the flow evolution from a particular initial condition, and considered a wide range of cases, from APG TBLs to the sink flow. More recently, a simplified version of the framework developed by \cite{Perry:2002cg} was used by \cite{marusic_et_al_jfm_2015} to evaluate the downstream evolution of ZPG TBLs from various initial conditions, in particular to assess the convergence towards a canonical ZPG configuration from different tripping conditions. 

The focus of this study is on near-equilibrium APG TBLs, and more precisely on the assessment of history effects on the boundary-layer development. To that end, we consider the Clauser pressure-gradient parameter $\beta= \delta^{*} / \tau_{w} {\rm d}P_{\infty} / {\rm d}x$, where $\delta^{*}$ is the displacement thickness, $\tau_{w}$ is the wall-shear stress and $P_{\infty}$ is the free-stream pressure, to quantify the pressure-gradient magnitude and evaluate the evolution of flat-plate TBLs under various $\beta(x)$ distributions. For this purpose, well-resolved large-eddy simulations (LES) of turbulent boundary layers with various APG conditions were carried out, and their results were compared with other available databases as described below.

\section{Numerical method and databases} \label{numerical_section}
The downstream evolution of TBLs subjected to adverse pressure gradients was studied by means of well-resolved large-eddy simulations (LESs). The pressure gradient was imposed through the variation of the free-stream velocity at the top of the domain, which was defined following the near-equilibrium definition by Townsend (1956), {\it i.e.}, $U_\infty(x)=C(x-x_0)^m$. We used the code SIMSON (Chevalier {\it et al.}, 2007), which is based on a fully-spectral method with Fourier discretisation in streamwise and spanwise directions and on the Chebyshev-tau method in the wall-normal direction. Using the approximate deconvolution relaxation-term model  as a sub-grid scale model as in Eitel-Amor {\it et al.} (2014), the resolution was chosen as $\Delta x^+=21.5$, $y^+_{{\rm max}}=13.9$ and $\Delta z^+=9.2$ (where $x$, $y$ and $z$ denote streamwise, wall-normal and spanwise coordinates, respectively), with 12 points below $y^+=10$.  At the wall a no-slip condition was imposed, while at the upper boundary a Neumann condition was applied. %Note that in particular, we set $\partial u/\partial y=\partial w /\partial y=0$, and the wall-normal velocity gradient is obtained from continuity as $\partial u / \partial y =-\partial U_{\infty} / \partial x$, therefore the freestream velocity distribution is imposed through the Neumann boundary condition.  
%In order to ensure periodic boundary conditions in the streamwise direction a fringe region was considered upstream of the outlet, which forces the fully-turbulent flow back to the laminar Blasius inlet profile. 
%The simulation-length scales are based on the displacement thickness of the laminar inlet profile. 
%Although technically this choice of boundary conditions does not lead to zero spanwise vorticity $\omega_{z}$ in the freestream, we have observed that $\omega_{z}$ is approximately zero well beyond the boundary layer edge (for $y>2.5 \delta_{99}$, where $\delta_{99}$ is the $99\%$ boundary layer thickness), and only very close to the top boundary increases slightly to match the condition $\partial u / \partial y=0$ Bobke {\it et al.} (2016).

Different near-equilibrium boundary layers were investigated by varying the virtual origin $x_0$ and the power-law exponent $m$ as listed in Table~\ref{tab:11}. The pressure gradients in those TBLs are of a different magnitude, and exhibit various streamwise developments. The resulting pressure gradient parameter $\beta$ decreases over the streamwise direction in the cases $m13$, $m16$ and $m18$, whereas $\beta$ remains constant over streamwise distances of $37 \overline{\delta}_{99}$ and $28 \overline{\delta}_{99}$ in the cases $b1$ and $b2$, respectively. Note that $\delta_{99}$ is the $99\%$ boundary-layer thickness averaged over the region where $\beta$ is observed to remain constant, and $\delta_{99}$ was determined by means of the method developed by Vinuesa {\it et al.} (2016). Further details regarding the numerical setup of cases $m13$, $m16$ and $m18$ are given by Bobke {\it et al.} (2016). In addition to the five flat-plate APG cases discussed above, in the present study we also consider the TBL developing over the suction side of a NACA4412 wing section at $Re_{c}=400,000$ (where $Re_{c}$ is the Reynolds number based on freestream velocity $U_{\infty}$ and chord length $c$) by Hosseini {{\it et al.} (2016), and the ZPG TBL data by Schlatter {\it et al.} (2009), as shown in Table \ref{tab:11}. The idea is that the TBL developing on the suction side of the wing is subjected to a progressively stronger APG  (contrary to the flat-plate APG TBLs, in which case they are either constant or mildly relaxing), and therefore exhibits a very interesting $\beta(x)$ distribution, which will be compared with the near-equilibrium cases developing over the flat plate. Direct numerical simulation (DNS) was considered for the wing case, and the spectral-element code Nek5000 (Fischer {\it et al.}, 2008) was employed, as discussed in detail in Hosseini {\it et al.} (2016). The DNS of ZPG TBL by Schlatter {\it et al.} (2009) is considered to provide a baseline case, with respect to which pressure-gradient effects can be assessed.

\begin{table}
\begin{center}
%{\footnotesize
\begin{tabular}{c c c c}
\hline
Case & Reynolds number range  & $\beta$ & Color code\\
\hline
$m13$  &$ 700 < Re_\theta < 3515$ & $[0.86;1.49]$ & {\color{dgreen}\solid}\\
$m16$  & $710 < Re_\theta < 4000$ & $[1.55;2.55]$ & \color{blue}{\solid}\\
$m18$   &$710 < Re_\theta < 4320$& $[2.15;4.07]$ & \color{violet}{\solid}\\
$b1$  &$670 < Re_\theta < 3360$ & $1$ & \color{orange}{\solid} \\
$b2$  &$685 < Re_\theta < 4000$ & $2$ & \color{brown}{\solid}\\
Wing  &$260 < Re_\theta < 2800$  & $[0;85]$ & \color{red}{\solid}\\
ZPG  &$670 < Re_\theta < 2500$ & $0$ & \color{black}{\solid}\\
\hline
\end{tabular}
\caption{List of datasets used in the present paper, including their momentum-loss Reynolds number range, power-law exponent, Clauser pressure-gradient parameter and color code used throughout the remainder of the paper. The setup of cases $m13$, $m16$ and $m18$ is reported in detail by Bobke {\it et al.} (2016); the Wing case is described by Hosseini {\it et al.} (2016), and the ZPG database is reported by Schlatter {\it et al.} (2009).}
\label{tab:11} 
\end{center}
\end{table}

\section{Effect of history on turbulence statistics} \label{statistics_section}
We first report the results of five near-equilibrium APG TBLs on flat plates, defined by different power-law exponents and virtual origins. As stated in $\S$1, the state of the boundary layer will not depend on the particular value of $\beta$ at a certain position, but on its development history, {\it i.e.}, on $\beta(x)$. While $\beta$ decreases over the streamwise direction in the cases denoted with $m$ ($m13$, $m16$, $m18$), $\beta$ remains constant for the two cases denoted with $b$ ($b1$, $b2$). Let us recall that although $\beta$ is not constant with $x$ in the $m$ cases, these TBLs are in near-equilibrium due to the fact that the $U_{\infty}$ distribution is prescribed by a power law as defined by Townsend (1956) and Mellor and Gibson (1966). Regarding the cases with constant $\beta$, not only are they in near-equilibrium, but they also allow a better characterization of Reynolds-number effects in a certain pressure-gradient configuration. Note that the ZPG TBL flow essentially corresponds to a constant $\beta=0$ configuration. In Figure~\ref{Fig:1} we show the streamwise evolution of $\beta$, as a function of the friction Reynolds number $Re_{\tau}$ and the streamwise component $x$, for the various flat-plate cases as well as for the TBL on the suction side of a wing described in $\S$2. For the flat-plate cases the inflow laminar displacement thickness $\delta^{*}_{0}$ is used to nondimensionalise $x$, whereas for the case of the wing the displacement thickness at $x/c=0.15$, where the flow is post-transitional, is considered. In order to evaluate the impact of different $\beta(x)$ distributions on the local state of the APG TBL, we select three cases in which we have the same $\beta$ and $Re_{\tau}$, but a different history of $\beta$. As highlighted with black dots in Figure \ref{Fig:1}, the first selected case is with $\beta=1.4$ and $Re_{\tau}=340$, obtained from the wing (which starts from very low values of $\beta$ and exhibits approximately exponential growth with $x$) and from the flat-plate case $m13$ (in which a decreasing trend in $\beta$, starting from higher values, is observed). The second case exhibits a slightly higher friction Reynolds number $Re_\tau=367$, at a stronger adverse pressure gradient $\beta=2.9$, and in this case also the wing is selected (with the exponentially increasing $\beta(x)$), together with the flat-plate APG case $m18$, which at that point exhibits a slightly increasing trend in $\beta$. The third case highlighted in Figure \ref{Fig:1} involves the two flat-plate APG TBLs $m16$ and $b2$, at a higher Reynolds number of $Re_\tau=762$, and with $\beta=2.0$. Note that in this particular case both boundary layers are in near-equilibrium, and that the $b2$ configuration exhibits a constant value of $\beta=2$ starting at $x \simeq 1000$, whereas in the $m16$ case the $\beta$ curve shows a decreasing trend. 
\begin{figure}
\begin{center}
\includegraphics*[width=0.9\linewidth]{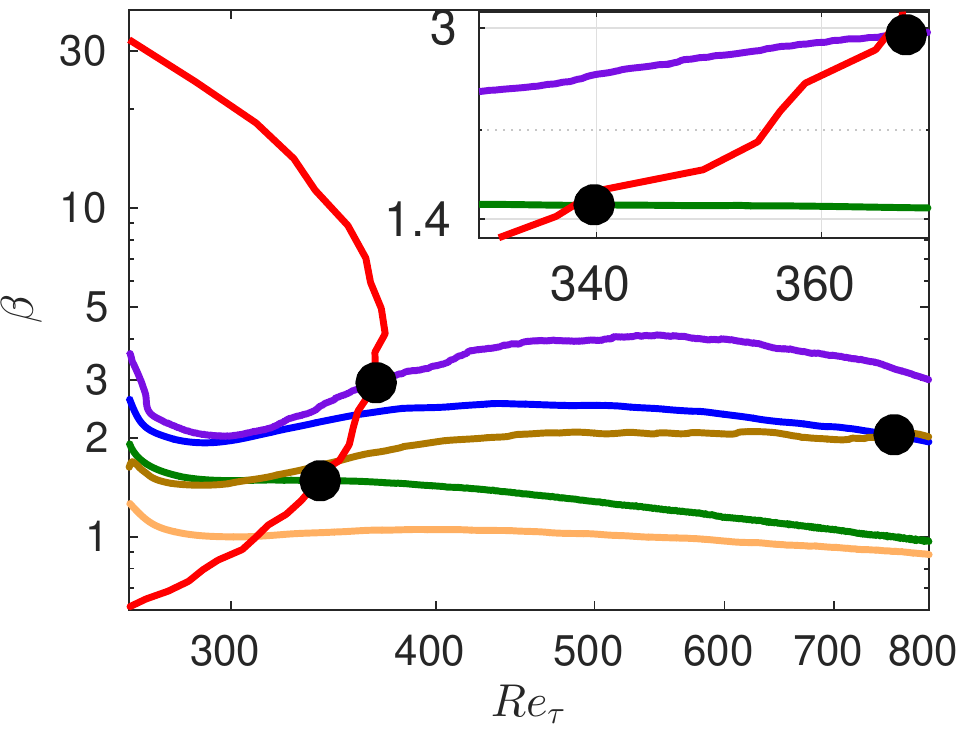}\put(-143,45){I}\put(-130,72){II}\put(-20,55){III}
\caption{Clauser pressure-gradient parameter $\beta$ as function of $Re_\tau$ for the following cases: boundary layer developing on the suction side of a wing (Hosseini {\it et al.}, 2016): red, and over a flat plate for non-constant $\beta$-cases ($m=-0.13$: green, $m=-0.16$: blue, $m=-0.18$: purple) and constant $\beta$-cases ($\beta=1$: orange, $\beta=2$: brown). Inset and black dots indicate the matched $\beta-Re_\tau$ values that will be considered in the remainder of the paper. }
\label{Fig:1}
\end{center}
\end{figure}

In Figure \ref{Fig:2} we show the inner-scaled mean flow for the various comparisons discussed above, as well as selected components of the Reynolds-stress tensor. The first two important observations from this figure are: although the three comparisons are at the same $\beta$ and $Re_{\tau}$, the turbulence statistics in the outer layer are essentially different among the cases, while they agree in the viscous region. This highlights the significant impact of history effects on the state of the outer layer of a turbulent boundary layer. Focusing on Figure \ref{Fig:2}(a), we can observe the general effect of a moderate APG with $\beta=1.4$ on the boundary layers, compared with the equivalent ZPG case: the APG TBLs exhibit a steeper logarithmic region, and a more prominent wake than the ZPG, associated with stronger energetic structures in the outer region, as also observed by Monty {\it et al.} (2011) and Vinuesa {\it et al.} (2014). With respect to the differences between the two APG cases, it is important to recall that the profiles on the suction side of the wing were obtained by means of DNS, whereas the flat-plate boundary layers are based on LES. This could be the reason for the subtle discrepancies between both profiles in the buffer region, since as documented by Eitel-Amor {\it et al.} (2014) the coarser resolution used in the LES produces slightly lower mean velocities in this region of the boundary layer. Nevertheless, the effect of the LES is negligible in the outer region, and therefore the differences observed in the wake of the two APG cases can be attributed to their particular streamwise evolution. Monty {\it et al.} (2011) showed that the APG energizes the largest turbulent structures in the outer flow, leading to the more prominent wake, as well as to the outer peak in the streamwise velocity fluctuation profile. As noticeable in Figure \ref{Fig:1}, the $\beta(x)$ curve from the $m13$ case exhibits values above $1.4$ from the start of the pressure-gradient region, whereas in the wing the initial $\beta$ values are close to zero, and they only reach the value 1.4 after a certain streamwise development. Therefore, in the $m13$ case the outer flow was subjected to a stronger APG throughout its streamwise development, and therefore the larger structures received much more energy from the APG. As a consequence, and although at $Re_{\tau}=340$ the wing and the flat-plate boundary layers have the same value of $\beta=1.4$, the accumulated $\beta(x)$ effect leads to a stronger impact of the APG in the $m13$ case. The Reynolds-stress tensor components are analyzed for this case in Figure \ref{Fig:2}(b), where again the most characteristic features of APG TBLs can be observed in comparison with the ZPG case (Monty {\it et al.}, 2011): the streamwise velocity fluctuation profile develops an outer peak, a consequence of the energizing of the large-scale motions, which also produces an increase of the near-wall peak due to the modulation of the near-wall region by the outer flow. Note that the location of this inner peak, $y^{+} \simeq 15$, is essentially unaffected by the APG. The wall-normal and spanwise velocity fluctuations, as well as the Reynolds shear-stress profile, exhibit a more prominent outer region compared with the ZPG due to the effect of the APG on the outer flow. Regarding the characteristics of the two APG cases, the first noticeable feature is the fact that the value of the inner peak appears to be approximately the same in the two cases, whereas the $m13$ case exhibits a stronger outer peak. The larger value of the outer peak can be explained, as well as the more prominent wake, by the fact that the flat-plate case was exposed to a higher accumulated $\beta(x)$, therefore the large-scale motions in the outer flow were effectively more energized than those in the wing. Nevertheless, it would be expected that the inner peak would also be larger in the $m13$ case, due to the modulation effect mentioned above. A possible explanation for this apparent contradiction lies in the use of LES for the $m13$, which as also mentioned above does not have a noticeable effect on the outer region. Interestingly, the outer-region wall-normal and spanwise fluctuations are also stronger in the $m13$ case than in the wing, although the Reynolds shear-stress profiles exhibit values slightly larger in the wing. An alternative explanation might be related to the different upstream histories of the boundary layers exposed to nearly the same $\beta$ parameter: the boundary layer on the wing increases in terms of the strength of the APG along its downstream evolution, while the TBL in the  $m=-0.13$ case stems from a stronger APG that relaxes in terms of $\beta$. Whereas the inner layer adapts quickly to the imposed pressure gradient, the outer layer inherits the different upstream histories further downstream, thereby yielding matched inner-layer turbulence statistics at the same $\beta$-value, while the outer layer exhibits amplitudes that are rather representative of the respective $\beta$ at a more upstream station, {\it i.e.}, a higher and lower $\beta$ value for the $m=-0.13$ and wing, respectively.

The second comparison is also between a flat-plate APG and the boundary layer developing on the suction side of the wing, this time at $\beta=2.9$ and $Re_{\tau}=367$. In Figure \ref{Fig:2}(c) the effect of a stronger APG on the mean flow can be observed in comparison with the ZPG, more precisely, the wake region is significantly stronger (a fact associated with much lower skin friction and the lifting up of the boundary layer by the action of the APG), and the incipient log region is steeper. Also in this case, the $\beta(x)$ history from the flat-plate case ($m18$) leads to higher accumulated effect of the APG in comparison with the wing. In particular, the $m18$ case exhibits values of $\beta$ starting around 2 (at the beginning of the APG region), and increasing up to the value of around 2.9 where the comparison with the wing is performed. On the other hand, in the wing the initial values are around zero and rise quickly up to the value of $2.9$, but the accumulated APG effect is significantly inferior to that of the flat-plate case. This is again manifested in the more prominent wake region from the $m18$ configuration, due to the fact that the most energetic structures in the outer flow have been exposed to a stronger APG throughout the boundary layer development. Interestingly, the discrepancy in the logarithmic and buffer regions is larger in this case than what was observed in Figure \ref{Fig:2}(a), at a lower $\beta$. Note that the lower velocities in the buffer region with stronger localized APGs have already been reported by Monty {\it et al.} (2011), and therefore it is plausible that in this case they could be caused by the different $\beta(x)$ from the two cases. Figure \ref{Fig:2}(d) further supports the fact that the accumulated $\beta(x)$ in the $m18$ case leads to a much more energetic outer region compared with the wing, although the local magnitude of $\beta$ and the Reynolds number are the same for the compared profiles. The outer peak in the streamwise fluctuations is significantly larger in the flat-plate case, and the differences in the outer region are also noticeable in the other two fluctuation components, as well as in the Reynolds shear stress (as opposed to what was observed in the lower $\beta$ case described above). Interestingly, also in this case the inner peak in the streamwise fluctuations from the two APG TBLs exhibits approximately the same value, despite the large difference in the outer region. The attenuation effect of the LES reported by Eitel-Amor {\it et al.} (2014) is also around $4\%$ at this $Re_{\tau}$ in ZPG TBLs, therefore it can also be argued that the inner peak would be marginally larger in the $m18$ than in the wing if a DNS had been performed. %\rev{While in this case both APGs inherit the history of  a lower $\beta$, the changes in $\beta$ in case of the wing are more drastic thereby giving the outer layer less development length to adapt itself to the local APG condition.}

\begin{figure}[!t]
\begin{center}
\includegraphics*[width=0.49\linewidth]{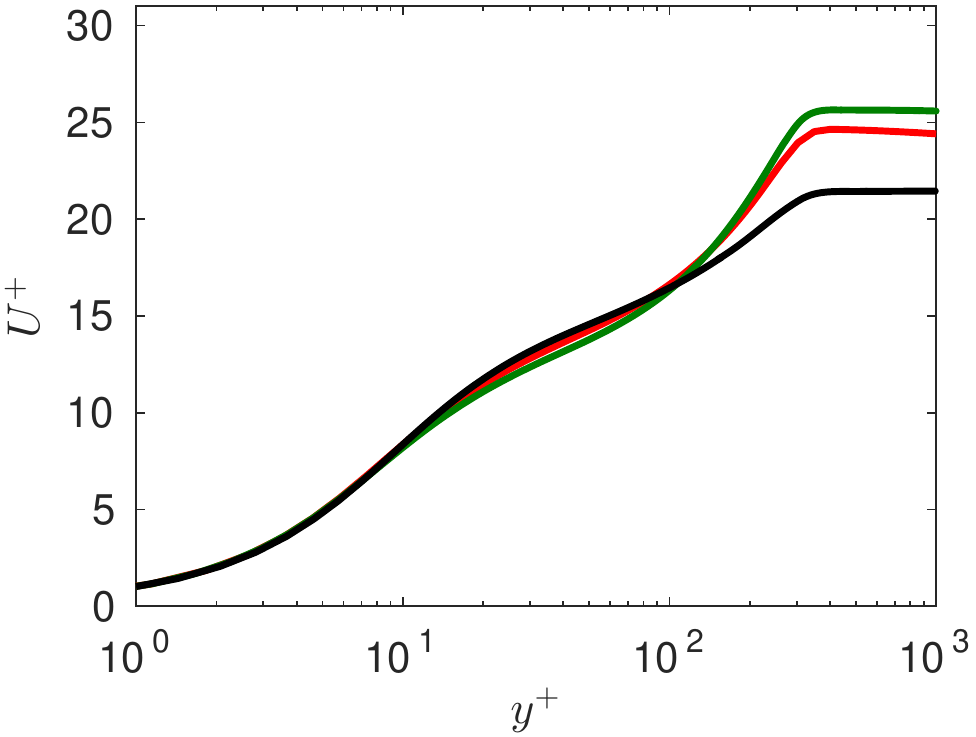}\put(-98,76){$(a)$}\hfill
\includegraphics*[width=0.49\linewidth]{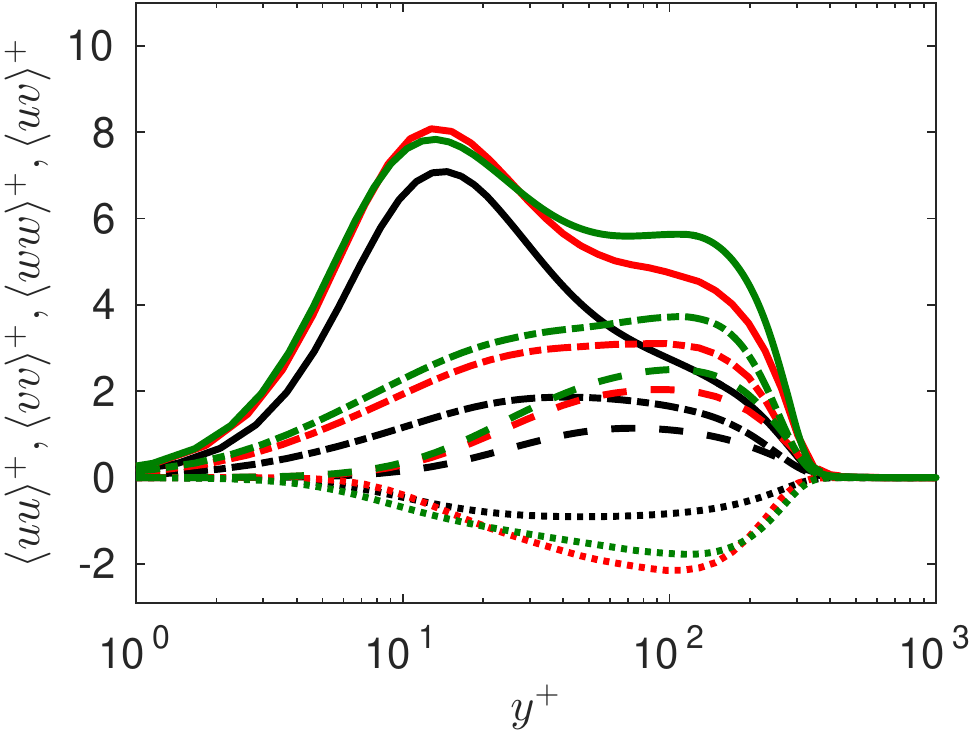}\put(-98,76){$(b)$}\\
\includegraphics*[width=0.49\linewidth]{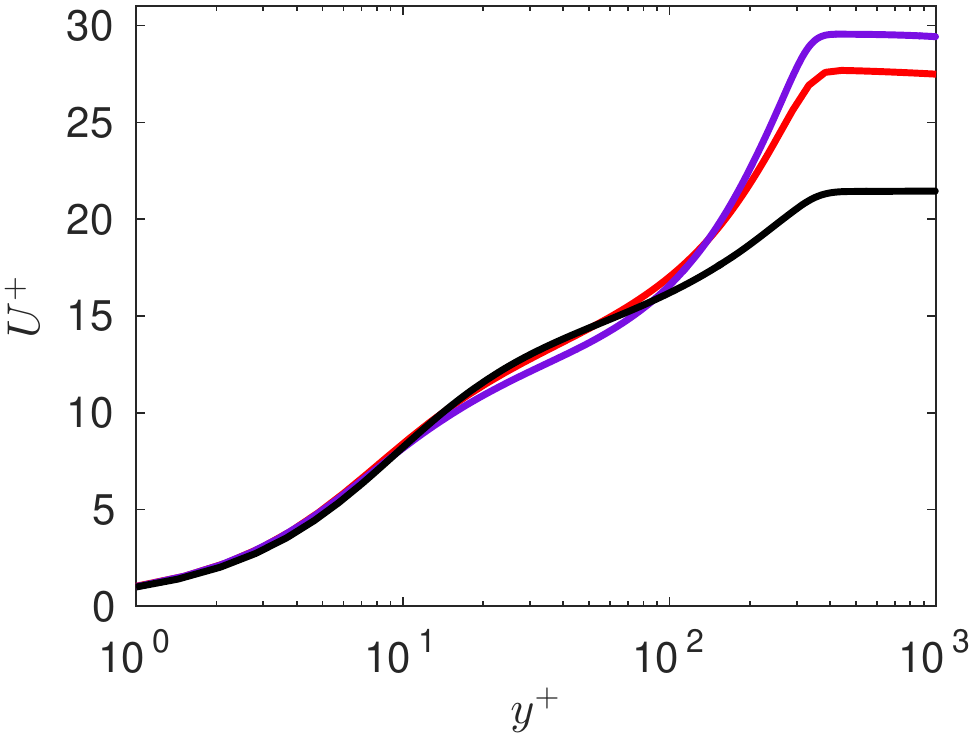}\put(-98,76){$(c)$}\hfill
\includegraphics*[width=0.49\linewidth]{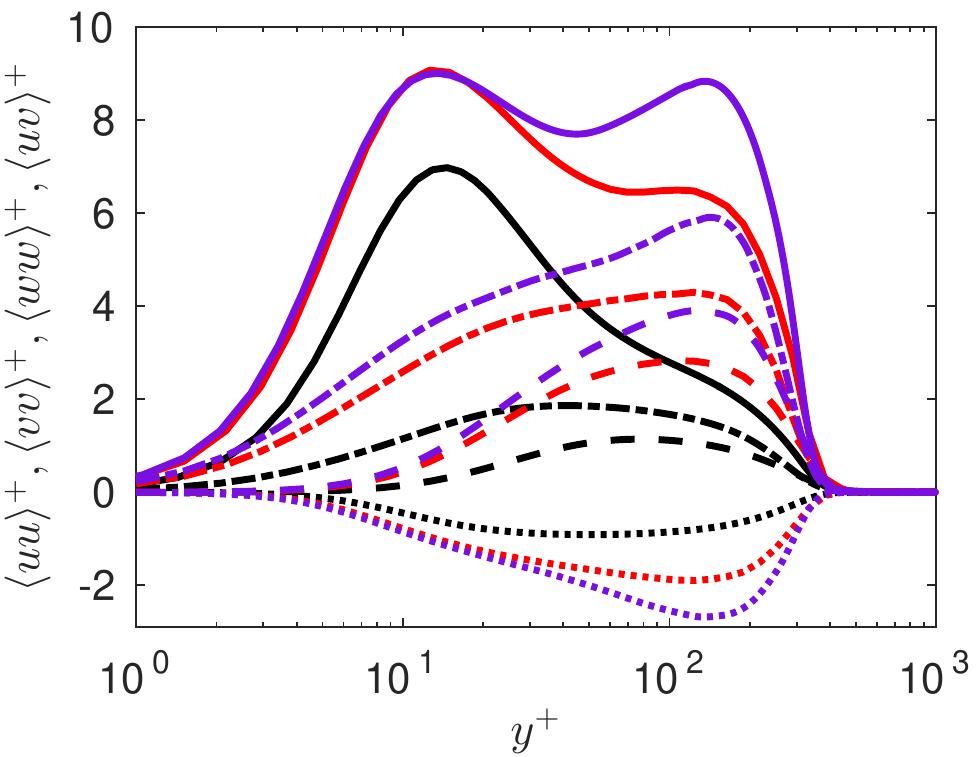}\put(-98,76){$(d)$}\\
\includegraphics*[width=0.49\linewidth]{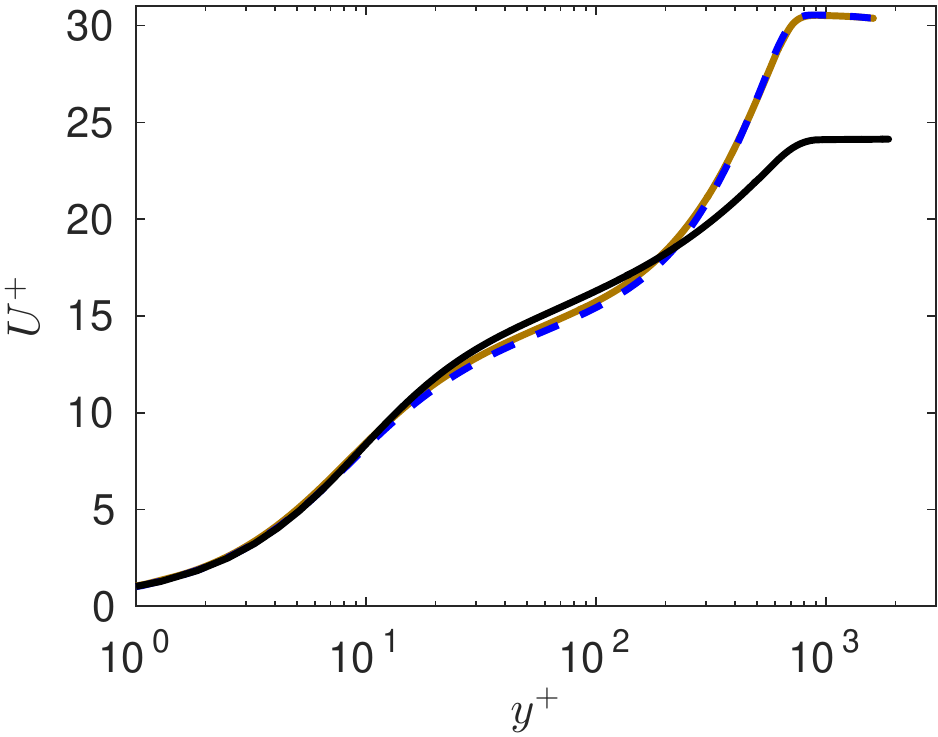}\put(-98,80){$(e)$}\hfill
\includegraphics*[width=0.49\linewidth]{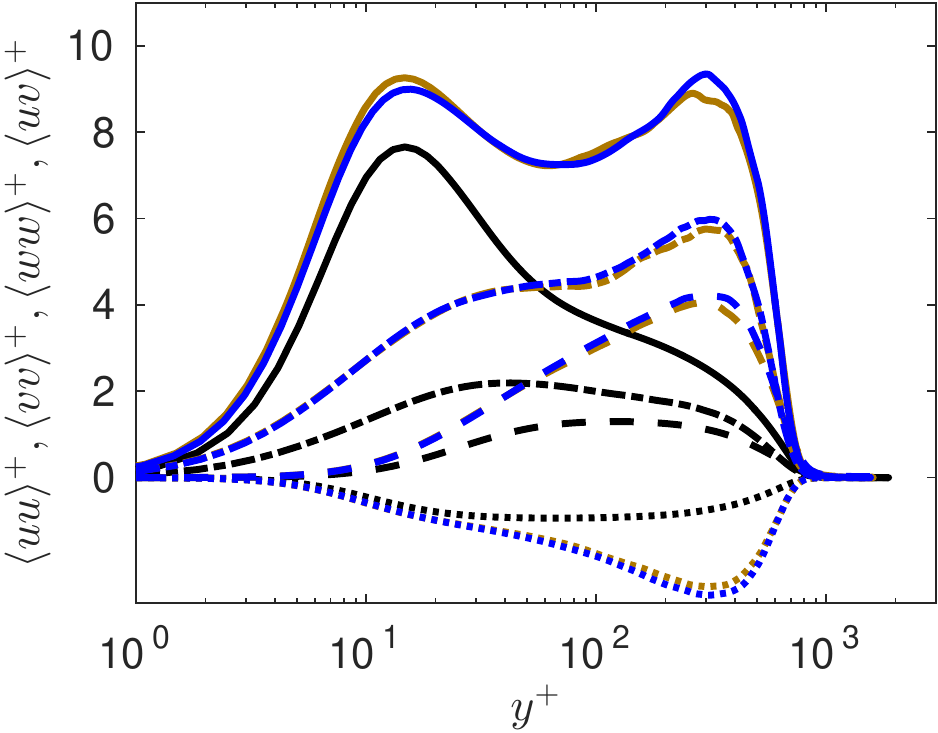}\put(-98,80){$(f)$}
 \caption{(a),(c),(e) Inner-scaled mean velocity profiles of the wing (red), $m=-0.13$ (green), $m=-0.16$ (blue), $m=-0.18$ (purple), $\beta=2$ (brown) and ZPG (black). (b),(d),(f) Variation of the inner-scaled Reynolds stress profiles: $\langle uu\rangle^+$ (solid), $\langle vv\rangle^+$ (dashed), $\langle ww\rangle^+$ (dot-dashed) and $\langle uv\rangle^+$ (dotted). (a),(b) I: $\beta=1.4$ and $Re_\tau=340$.  (c),(d) II: $\beta=2.9$ and $Re_\tau=367$. (e),(f) III: $\beta=2.0$ and $Re_\tau=762$.} 
\label{Fig:2}
\end{center}
\end{figure}

Finally, in Figure \ref{Fig:2}(e) we compare the mean flow from two flat-plate cases, one with a constant $\beta$ region ($b2$), and the other one with no constant $\beta$ ($m16$); both in near-equilibrium. In particular, the comparison is done at $\beta=2$ and at a higher friction Reynolds number of $Re_{\tau}=762$. The effect of the APG is also noticeable in this case, with the more prominent wake and lower velocities in the buffer region, in comparison with the ZPG. Note that the $U^{+}_{\infty}$ value from the flat-plate boundary layers is around 30, approximately the value obtained in the $m18$ case. Interestingly, this inner-scaled freestream velocity is obtained here with a lower $\beta$ (2 instead of 2.9), but higher $Re_{\tau}$ (762 instead of 367). This, together with the similarities between Figure \ref{Fig:2}(f) and d) in terms of inner and outer peaks of the streamwise velocity fluctuations, suggests certain connections between Reynolds-number and pressure-gradient effects. More precisely, a low-$Re$ APG TBL may exhibit features of a high-$Re$ ZPG TBL, if the magnitude of the APG is strong enough. This also points towards certain connections between the energizing mechanisms in the outer flow present at higher Reynolds numbers and with stronger APGs. Focusing on Figure \ref{Fig:2}(e), it is interesting to note that the two flat-plate cases exhibit very good agreement in their mean flow profiles, although their streamwise developments are different. Nevertheless, Figure \ref{Fig:1}(b) shows that although the $m16$ exhibits a decreasing trend in $\beta(x)$, and in the $b2$ a region of constant $\beta$ is observed, from $x \simeq 1500$ to around $2000$ (location where the comparison is done), the two curves converge, and the relative differences between the two curves are below $15\%$. Hence, both APGs share a similar upstream history for about 6.5 local boundary-layer thicknesses. Regarding the components of the Reynolds-stress tensor shown in Figure \ref{Fig:2}(f), first of all the pressure gradient effects (combined with the moderate $Re_{\tau}$ of 762) lead to significantly more energized components in the outer region compared with the ZPG, as well as a larger near-wall peak in the streamwise component. Interestingly, in this case the outer peak of the streamwise velocity fluctuations is slightly larger than the inner peak; a phenomenon that suggests the development of a different energy distribution throughout the boundary layer, compared with that of moderately high ZPG TBLs. Such an overtaking of the inner peak by an outer peak has for instance been predicted by the diagnostic profile as shown by Alfredsson {\it et al.} (2012), although there, the outer peak resided within the overlap region, which is not the case for strong APGs. The other significant observation is the fact that the two flat-plate APG boundary layers exhibit very good agreement in all the components of the Reynolds-stress tensor, again highlighting the convergence of the two boundary layers towards the same state. These results suggest that, in this particular configuration with a moderately changing $\beta$, a streamwise distance of around $x/\delta^{*}_{0} \simeq 500$ (where $\delta^{*}_{0}$ is the displacement thickness of the inflow laminar boundary layer), corresponding to $6.5 \delta_{99}$, may be sufficient for the APG TBL to become independent of its initial downstream development, and converge towards a certain state uniquely characterized by the $\beta$ and $Re_{\tau}$ values.

\section{Assessment of alternative scaling laws}

Due to the significant impact of history effects on the local flow features as discussed above, in this work we aim at characterizing configurations with values of $\beta$ constant over a significant portion of the domain. As observed by Mellor and Gibson (1966), the constant $\beta$ configuration is a particular case of near-equilibrium TBL, and therefore the $U_{\infty}(x)$ is also defined by a power law with particular choices of $x_{0}$ and $m$. A detailed characterization of constant $\beta$ cases will ultimately allow to assess pressure-gradient effects with progressively more complex history effects, given by the particular $\beta(x)$ distribution. In the present work we obtained a configuration exhibiting a constant value of $\beta=1$ in the range $500 < x < 2300$, and another one with a constant value of $\beta \simeq 2$ in the range $1000 < x < 2300$. In Figure \ref{Fig:8} we show a schematic representation of the constant $\beta=1$ region, in comparison with the one obtained in the recent work by Kitsios {\it et al.} (2015), also for a constant $\beta=1$ case. Note that although Kitsios {\it et al.} (2015) explored higher Reynolds numbers than the ones considered here, the range over which $\beta$ is constant is 1.6 times larger in the present simulation.

\begin{figure}
\begin{center}
\includegraphics*[width=0.8\linewidth]{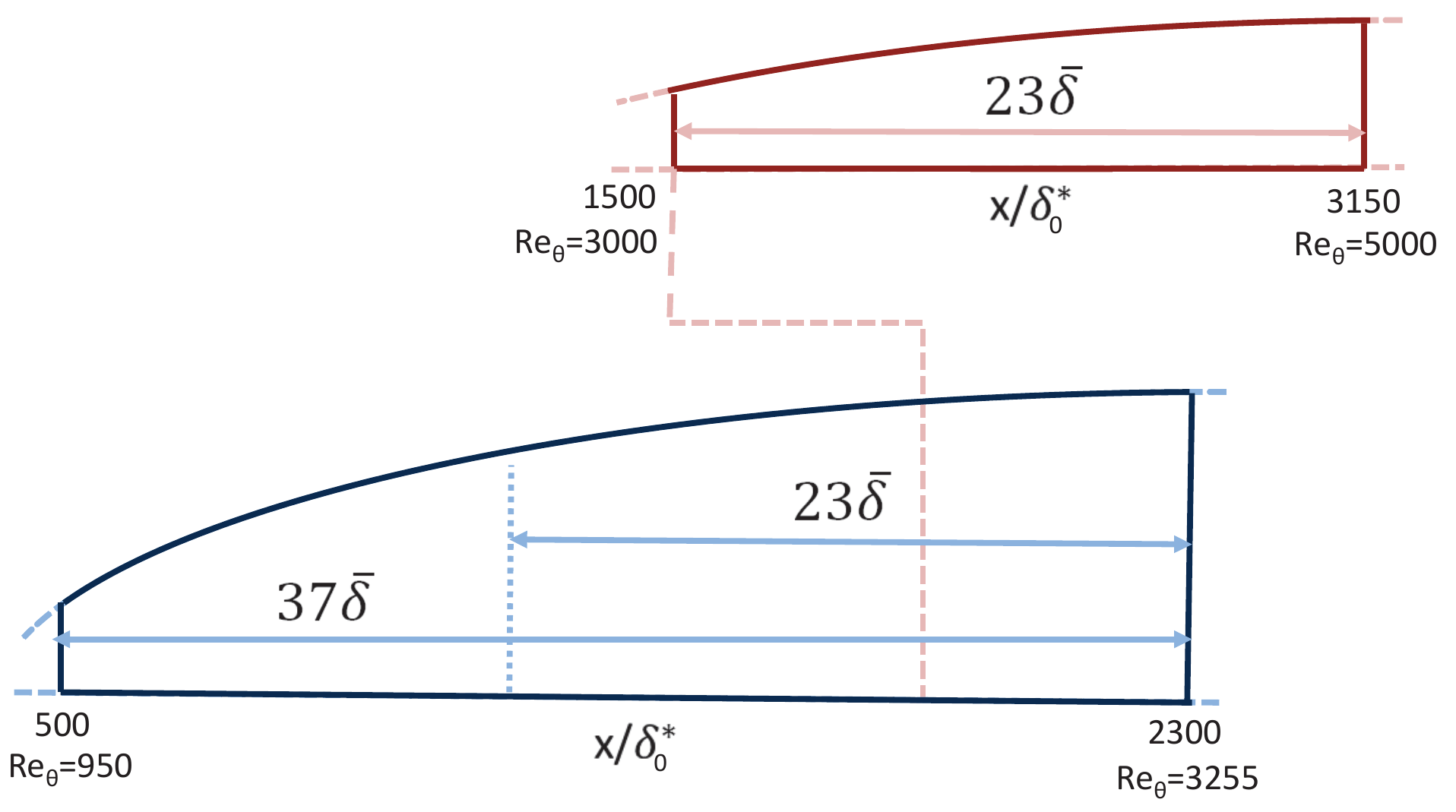}
\caption{(Blue) Sketch of the APG turbulent boundary layer showing the area where a constant value of $\beta=1$ was obtained, where $\delta^{*}_{0}$ is the displacement thickness of the laminar inflow boundary layer. (Red) Domain of interest with $\beta=1$ extracted from the study by Kitsios {\it et al.} (2015). The extent of the constant $\beta=1$ regions are shown in both cases normalized with the averaged boundary-layer thicknesses $\overline{\delta}$. The extent of the domain of interest from Kitsios {\it et al.} (2015) is also represented in our case.}   
\label{Fig:8}
\end{center}
\end{figure}

Figures \ref{Fig:91}(a) and (b) show the inner-scaled mean flow and velocity fluctuations corresponding to case $b1$, schematically discussed in Figure~\ref{Fig:8}. The profiles within the region of constant $\beta=1$ are highlighted in the two panels. The mean flow shows all characteristic features of APG TBLs, as discussed in $\S$3. Moreover, the velocity fluctuations develop an outer peak in all components, connected with the most energetic structures in the outer region. An alternative scaling for these quantities was considered by Kitsios {\it et al.} (2015) in their simulation, based on the displacement thickness $\delta^*$ and the local edge velocity $U_{e}$. They observed an apparent collapse of the mean flow and the fluctuations in their region of constant $\beta$, which as indicated in Figure \ref{Fig:8} corresponds to a streamwise distance of around $23$ integrated boundary-layer thicknesses $\overline{\delta}$. In Figures \ref{Fig:91}(c) and (d) we apply the same scaling to our data, and we do not observe such a collapse in any of the investigated quantities in our constant $\beta$ region, which spans a longer streamwise distance of $37 \overline{\delta}$. One possible explanation for this discrepancy could be that the scaling considered by Kitsios {\it et al.} (2015) does not lead to self-similarity, and since their constant $\beta$ region is shorter than ours  and their $Re$-range spans only $20\%$ of a decade, their streamwise development would be insufficient to reveal this conclusion. The present data exhibits a clear $Re$ trend  (spanning $23\%$ of a decade in $Re_{\theta}$), which is furthermore extended through the higher $Re$ data by Kitsios {\it et al.} (2015). This would indeed be in agreement with Townsend (1956), since in principle the sink flow is the only flow that can be described from the wall to the free-stream in terms of a single similarity variable in $y$. These aspects are further explored by analyzing the constant $\beta=2$ case, over a streamwise distance of $28 \overline{\delta}$. A higher $Re_\theta$ range is reached in this case, which is more comparable to the one analyzed by Kitsios {\it et al.} (2015), albeit at a higher value of $\beta$. As seen from Figures~\ref{Fig:91}(e) and (f), the scaling by Kitsios {\it et al.} (2015) does not lead to self-similarity in this case either. Also here a clear $Re$ trend is noticed, supporting the statements presented above, and also the validity of the classic two-layer similarity, at least for the $\beta$ range under consideration.  
\begin{figure}
\begin{center}
\includegraphics*[width=0.49\linewidth]{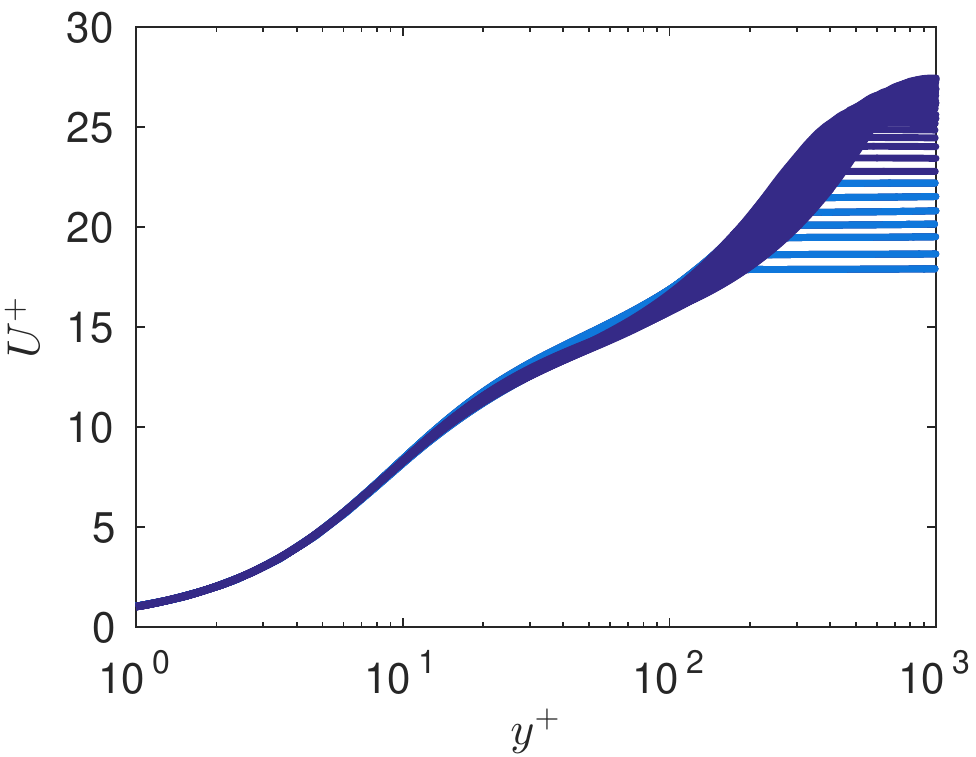}\put(-100,80){$(a)$}
\includegraphics*[width=0.49\linewidth]{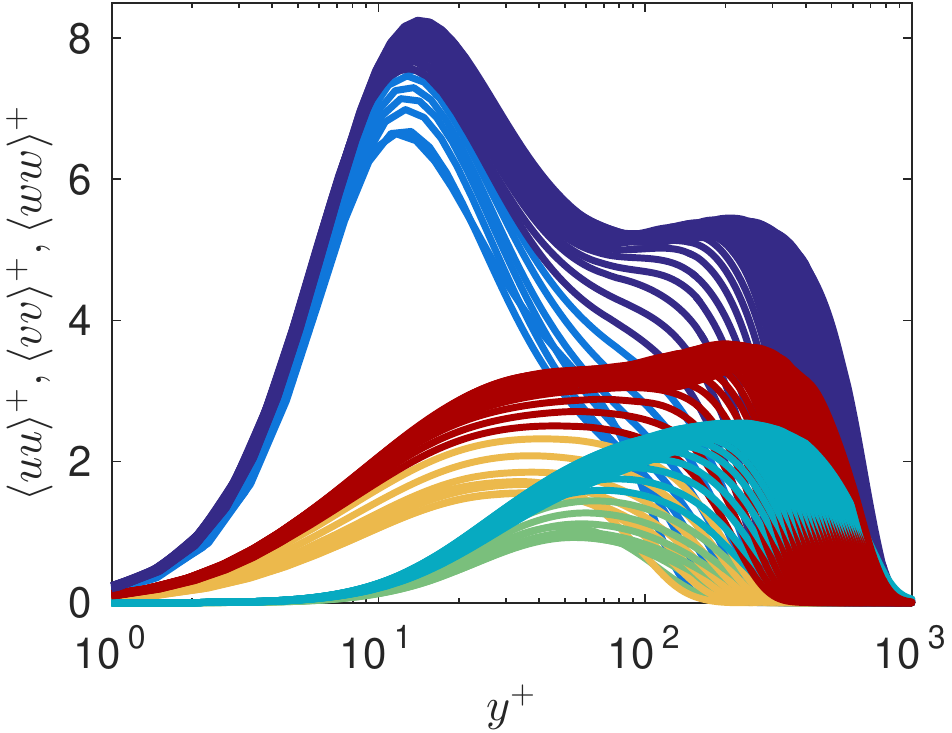}\put(-20,80){$(b)$} \\
\includegraphics*[width=0.49\linewidth]{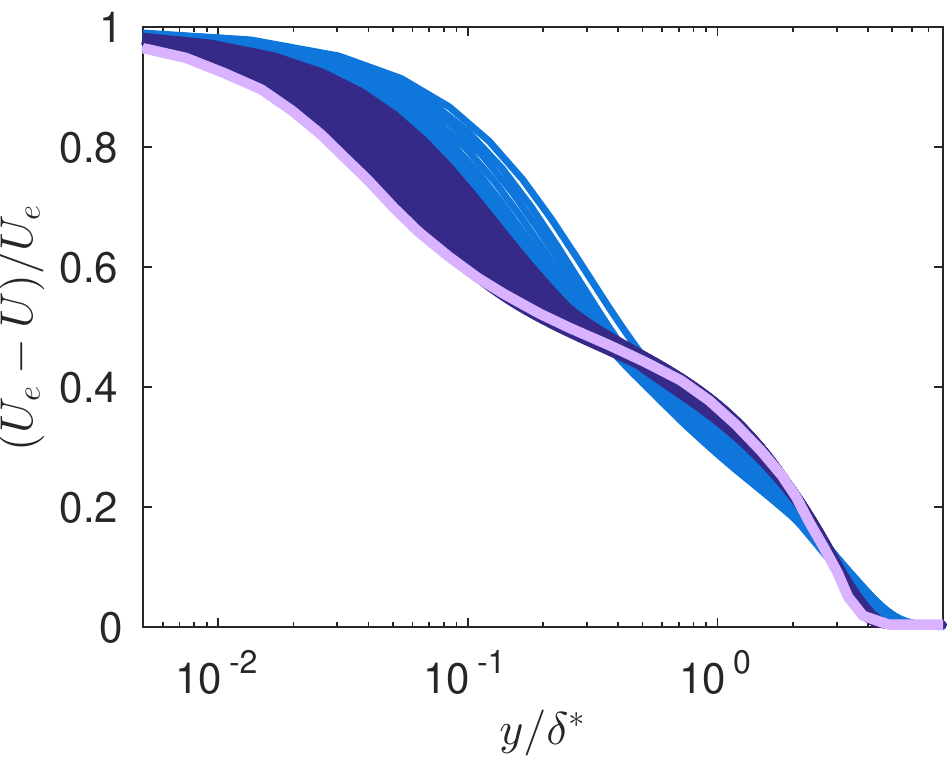}\put(-20,80){$(c)$}
\includegraphics*[width=0.49\linewidth]{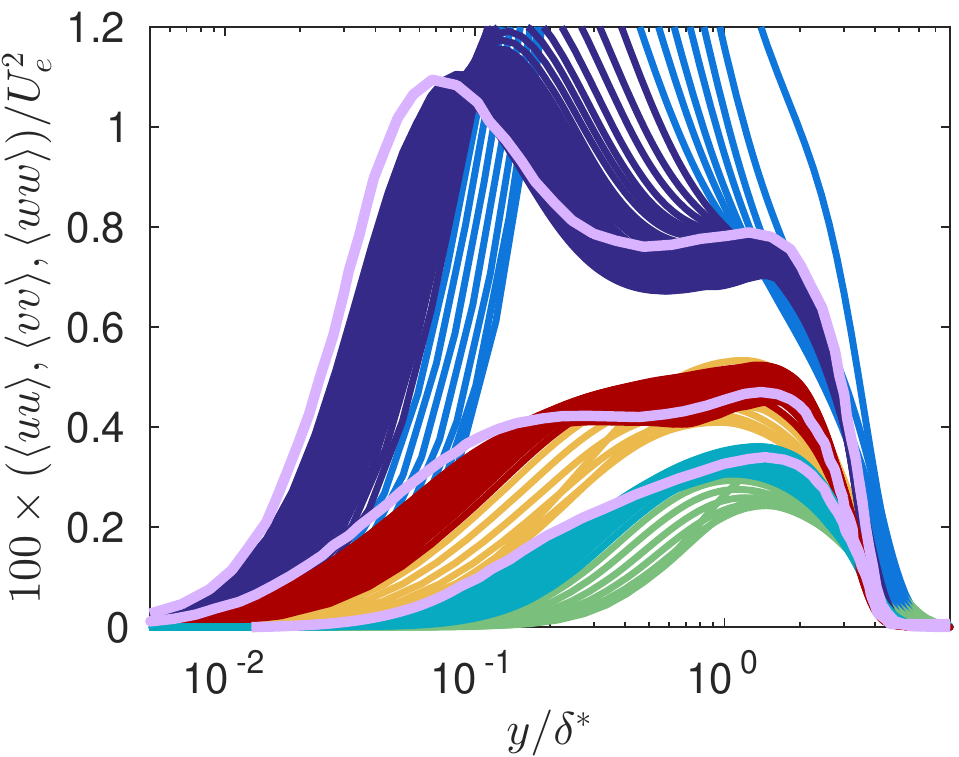}\put(-15,80){$(d)$} \\
\includegraphics*[width=0.49\linewidth]{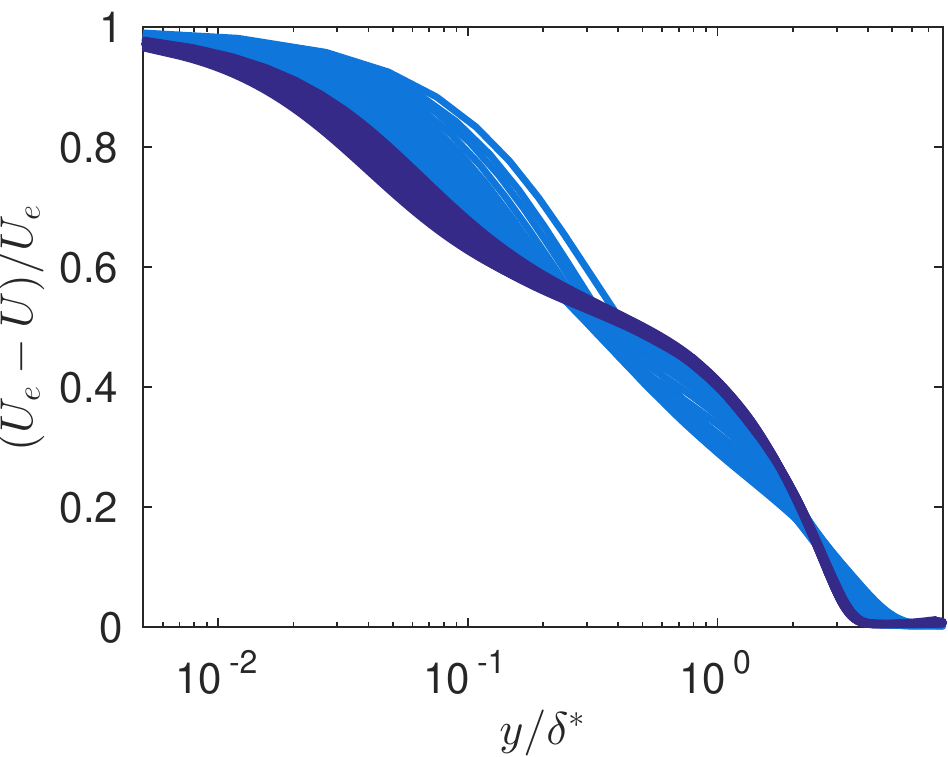}\put(-20,80){$(e)$}
\includegraphics*[width=0.49\linewidth]{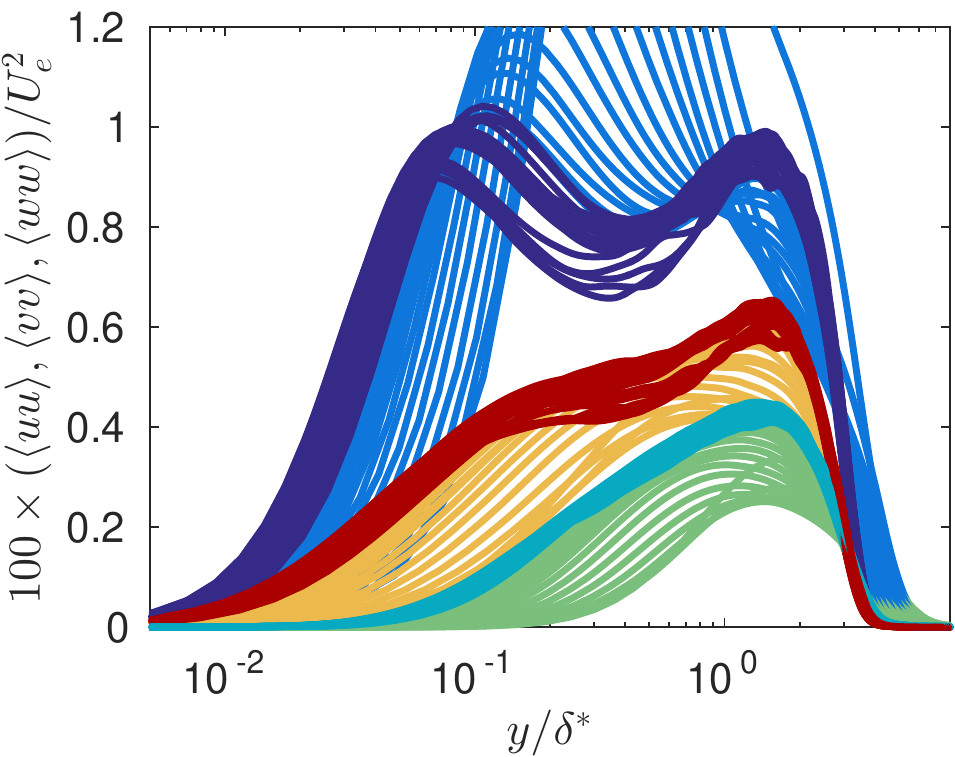}\put(-15,80){$(f)$}
\caption{Case $\beta=1$: Mean velocity profiles ($31$ positions in the range of $100<x<2300$) non-dimensionalised by (a) $u_\tau$ and $l^*=\nu/u_\tau$, (c) $U_e$ and $\delta^*$. Dark blue indicates the area of constant $\beta$ and light blue the non-constant $\beta$ region. Reynolds-stress profiles non-dimensionalised by (b) $u_\tau$ and $l^*=\nu/u_\tau$, (d) $U_e$ and $\delta^*$. Dark blue, turquoise, red denotes the constant $\beta$ region for $\langle uu \rangle$,$\langle vv \rangle$,$\langle ww \rangle$, respectively, and light blue, green, yellow indicate the non-constant $\beta$ region. Scaled profiles reported by Kitsios {\it et al.} (2015): purple. 
Case $\beta=2$: (e) Mean velocity profiles ($23$ positions in the range of $100<x<2300$). (f) Reynolds-stress profiles.}   
\label{Fig:91}
\end{center}
\end{figure}

\section{Conclusions}
The present study is focused on the history effects in turbulent near-equilibrium boundary layers with pressure gradients. After defining the {\it{near-equilibrium}} state according to Townsend (1956), large-eddy simulations were performed over a flat plate to assess the effect of different evolutions of the pressure-gradient parameter $\beta$. The adverse pressure gradient  was imposed by a varying free-stream velocity profile at the top of the domain, \emph{i.e.}, in the free-stream. Hereby constant and non-constant pressure distributions were achieved. With the constant pressure gradients, turbulent boundary layers at a certain state (due to the imposed pressure distribution), can be investigated over a wide range of Reynolds numbers. An interesting finding was obtained when comparing the mean and Reynolds stress profiles of the non-constant pressure and constant APG TBLs at matched  $\beta$ and $Re_\tau$. The non-constant $\beta$ case appears to converge towards the canonical state after a sufficiently long downstream length. For the conditions investigated in the present study, this length is $6.5\delta_{99}$.%This was also confirmed in the budget terms pertaining to the turbulent kinetic energy. 
 The history effects were studied not only in flat-plate TBLs, but also in the APG boundary layer developing over the suction side of a NACA4412 wing section. The large structures in the outer region were found to be less energetic on the suction side of the wing than in the flow over the flat plate for matched $\beta$ and $Re_\tau$. The structures were exposed to a lower PG over the streamwise direction (compared to the ones over the flat plate), resulting in a less pronounced wake region and a less intense outer region in the Reynolds stresses. %The energy budget also shows some of the reported effects of the APG, such as the higher values in the velocity-pressure-gradient correlation in the buffer and near-wall region or the extra wall dissipation. The comparison between the flow over a wing and the flat plate yield interesting results. The diffusion term is stronger close to the plate than over the surface of the wing at same $\beta$. This is due to the stronger wall dissipation resulting from the larger accumulated effect of the $\beta$ history. 
%One-dimensional spectral density maps indicate that the large-scale structures are energized in APG TBLs compared to ZPG TBLs at matched friction Reynolds numbers. The position of the near-wall spectral peak appears to be independent of the APG. This was also reported for high Reynolds number ZPG TBLs. 
A connection between PG TBLs and high-$Re$ ZPG TBLs might be able to be drawn, since the mechanisms, by which the large-scale motions are energised, in APG TBLs seem to share some similar features with those present in high-$Re$ ZPG TBLs.
Finally, we investigated the scaling proposed by Kitsios {\it et al.} (2015), in which $\delta^{*}$ and $U_{e}$ are considered as length and velocity scales. Our results show that this scaling does not lead to self-similar boundary layer profiles in the constant $\beta$ region. This conclusion is in agreement with Townsend, who showed that the sink-flow is the only boundary layer exhibiting self-similarity. Stronger streamwise constant pressure gradients at higher Reynolds numbers should be investigated in order to characterise cases closer to wind-tunnel experiments and general applications.

%%%%%%%%%% Insert here acknowledgments if necessary

\Acknowledgments
The simulations were performed on resources provided by the Swedish National Infrastructure for Computing (SNIC) at the PDC Center for High Performance Computing at KTH Stockholm. Financial support provided by the Knut and Alice Wallenberg Foundation and the Swedish Research Council (VR) is gratefully acknowledged.

\begin{References}

\item Townsend, A. A. (1956), The Structure of Turbulent Shear Flow, Cambridge Univ. Press, Cambridge, UK.
\item Marusic, I., Mckeon, B. J., Monkewitz, P. A., Nagib, H. M., Smits, A. J. and Sreenivasan, K. R. (2010), Wall-bounded turbulent flows at high Reynolds numbers: Recent advances and key issues, {\it Phys. Fluids}, Vol. 22, pp. 065103. 
\item Mellor, G. L. and Gibson, D. M. (1966), Equilibrium turbulent boundary layers, {\it J. Fluid Mech.}, Vol. 24, pp. 225--253. 
\item Schlatter, P., {\"O}rl{\"u}, R., Li, Q., Brethouwer, G., Fransson, J. H. M., Johansson, A. V., Alfredsson, P. H. and Henningson, D. S. (2009), Turbulent boundary layers up to ${Re}_\theta=$2500 studied through  simulation and experiment, {\it Phys. Fluids}, Vol. 21, pp. 051702.  
\item Bailey, S. C. C., Hultmark, M., Monty, J. P., Alfredsson, P. H., Chong, M. S., Duncan, R. D., Fransson, J. H. M., Hutchins, N., Marusic, I., Mckeon, B. J., Nagib, H. M., {\"O}rl{\"u}, R., Segalini, A., Smits, A. J. and Vinuesa, R. (2013), Obtaining accurate mean velocity measurements in high Reynolds number turbulent boundary layers using Pitot tubes, {\it J. Fluid Mech.}, Vol. 715, pp. 642--670.   
\item Chevalier, M., Schlatter, P., Lundbladh, A. and Henningson, D. S. (2007), SIMSON-- A pseudo-spectral solver for incompressible boundary layer flow, {\it Tech. Rep. TRITA-MEK 2007:07}, KTH Mechanics.
\item Eitel-Amor, G., {\"O}rl{\"u}, R. and Schlatter, P. (2014), Simulation and validation of a spatially evolving turbulent boundary layer up to $Re_{\theta}= 8300$, {\it Int. J. Heat Fluid Flow}, Vol. 47, pp. 57--69. 
\item Vinuesa, R., Bobke, A., {\"O}rl{\"u}, R. and Schlatter, P. (2016), On determining characteristic length scales in pressure-gradient turbulent boundary layers, {\it Phys. Fluids}, Vol. 28, pp. 055101. 
\item Bobke, A., Vinuesa, R., {\"O}rl{\"u}, R. and Schlatter, P. (2016), Large-eddy simulations of adverse pressure gradient turbulent boundary layers, {\it J. Phys.: Conf. Ser.}, Vol. 708, pp. 012012. 
\item Hosseini, S. M., Vinuesa, R., Schlatter, P., Hanifi, A. and Henningson, D. S. (2016), Direct numerical simulation of the flow around a wing section at moderate {R}eynolds number, {\it Int. J. Heat Fluid Flow}, doi:10.1016/j.ijheatfluidflow.2016.02.001 
\item Fischer, P. F., Lottes, J. W. and Kerkemeier, S. G. (2008), Nek5000: Open source spectral element CFD solver, Available at: \url{http://nek5000.mcs.anl.gov}
\item Monty, J. P., Harun, Z. and Marusic, I. (2011), A parametric study of adverse pressure gradient turbulent boundary layers, {\it Int. J. Heat Fluid Flow}, Vol. 32, pp. 575--585. 
\item Vinuesa, R., Rozier, P. H., Schlatter, P. and Nagib, H. M. (2014), Experiments and computations of localized pressure gradients with different history effects, {\it AIAA J.}, Vol. 55, pp. 368--384. 
\item Alfredsson, P. H., {\"O}rl{\"u}, R. and Segalini, A. (2012), A new formulation for the streamwise turbulence intensity distribution in wall-bounded turbulent flows, {\it Eur. J. Mech. B/Fluids}, Vol. 36, pp. 167--175.
\item Kitsios, V., Atkinson, C., Sillero, J. A., Borrell, G., Gungor, A., Jim\'enez, J. and Soria, J. (2015), Direct numerical simulation of a self-similar adverse pressure-gradient turbulent boundary layer, {\it Proc. 9th Intl Symp. on Turbulence and Shear Flow Phenomena}, Melbourne, Australia.

\end{References}
%\end{References}
%
%
\end{document}

%% file: symbollines.tex
\font\smallfont=cmsy10 at 10truept
\mathchardef\bigCircle="280D

\font\bigfont=cmsy10 at 14.4truept
\mathchardef\tiMes="2902        %

\font\Bigfont=cmsy10 at 17.28truept
\mathchardef\DiaMond="2A05        %
\mathchardef\cirCle="2A0E
\mathchardef\BigCircle="2A0D

\font\Bbigfont=cmsy10 at 24.88truept
\mathchardef\buLLet="2B0F

\def\bigCirc{\raise 0.3ex\hbox{$\bigCircle$}\nobreak$\,$}
\def\Bullet{\raise-0.35ex\hbox{$\buLLet$}\nobreak$\,$}

\def\triangledown{\raise 0.2em\hbox{$\bigtriangledown$}\nobreak$\,$}

\def\minisquare{\hbox{${\vcenter{
               \hrule height 0.3pt \kern-0.4pt
               \hbox{\vrule width  0.3pt height 3.0pt \kern 2.6pt
               \vrule width  0.3pt height 3.0pt} \kern-0.4pt
               \hrule height 0.3pt}}$}}
\def\ssquare{\raise 0.175ex\hbox{${\vcenter{
               \hrule height 0.5truept       \kern-0.25truept
               \hbox{\vrule width 0.5truept height 3.0truept \kern 2.75truept
                     \vrule width 0.5truept height 3.0truept} \kern-0.25truept
               \hrule height 0.5truept}}$}\nobreak$\,$}
\def\squarex{\raise 0.175ex\hbox{${\vcenter{
               \hrule height 0.8truept       \kern-1.80truept
          \hbox{\vrule width 0.8truept height 8.0truept \kern-1.95truept
                \raise 0.8truept\hbox{$\tiMes$}     \kern-6.70truept
                \vrule width 0.8truept height 8.0truept} \kern-0.80truept
               \hrule height 0.8truept}}$}\nobreak$\,$}

\def\sqbull{\raise0.175ex\hbox{\vrule height 1.4ex width 1.6ex depth 0.2ex}\nobreak$\,$}
\def\smsqbull{\raise0.175ex\hbox{\vrule height 0.8ex width 0.9ex depth 0.2ex}\nobreak$\,$}

\def\Diamondplus{${\vcenter{\vcenter{\DiaMond} \kern-10truept
                            \hbox{\vrule width .4truept}\kern -3truept
                            \hrule height .4truept}}$\nobreak$\,$}
\newcount\ndots

\def\drawline#1#2{\raise 2.5truept\vbox{\hrule width #1truept height #2truept}}
\def\moonspace#1{\hskip #1truept}

\def\Dashy{\drawline{4.00}{1.00}}     
\def\dashy{\drawline{4.00}{0.75}}     
\def\thindashy{\drawline{4.00}{0.25}}     
\def\dashyspace{\dashy\moonspace{2}}
\def\Dashyspace{\Dashy\moonspace{2}}
\def\thindashyspace{\thindashy\moonspace{2}}
\def\longdashy{\drawline{8.00}{0.75}} 
\def\thinlongdashy{\drawline{8.00}{0.25}} 
\def\longdashyspace{\longdashy\moonspace{2}}
\def\thinlongdashyspace{\thinlongdashy\moonspace{2}}

\def\solid{\drawline{24}{0.75}\nobreak$\,$}

\def\dashbox{\hbox{\dashyspace}}  
\def\Dashbox{\hbox{\Dashyspace}}  
\def\dashed{\hbox {\ndots=0 \loop\ifnum\ndots<3\advance\ndots by 1
        \dashbox\repeat\dashy}\nobreak$\,$}       
\def\Dashed{\hbox {\ndots=0 \loop\ifnum\ndots<3\advance\ndots by 1
        \Dashbox\repeat\Dashy}\nobreak$\,$}       
\def\thindashbox{\hbox{\thindashyspace}}  
\def\thindashed{\hbox {\ndots=0 \loop\ifnum\ndots<3\advance\ndots by 1
        \thindashbox\repeat\thindashy}\nobreak$\,$}       
\def\thindash{\hbox {\ndots=0 \loop\ifnum\ndots<3\advance\ndots by 1
        \thindashbox\repeat\thindashy}\nobreak$\,$}       
\def\longdashbox{\hbox{\longdashyspace}}  
\def\thinlongdashbox{\hbox{\thinlongdashyspace}}  
\def\longdash{\hbox {\ndots=0 \loop\ifnum\ndots<3\advance\ndots by 1
        \longdashbox\repeat\longdashy}\nobreak$\,$}       
\def\thinlongdash{\hbox {\ndots=0 \loop\ifnum\ndots<3\advance\ndots by 1
        \thinlongdashbox\repeat\thinlongdashy}\nobreak$\,$}       